\documentclass[ twocolumn,superscriptaddress, amsmath, showpacs,
tightenlines,twocolumn,twoside,prl]{revtex4}%
\usepackage{amsfonts}
\usepackage{amstext}
\usepackage{amsmath}
\usepackage{amssymb}
\usepackage{latexsym}
\usepackage{graphicx,epstopdf}
\usepackage{ulem}
\usepackage[colorlinks,linkcolor=blue,anchorcolor=blue,citecolor=blue]%
{hyperref}
\usepackage{times}
\usepackage{subfigure}
\usepackage{color}
\usepackage{dcolumn}
\usepackage{graphicx}%
\providecommand{\U}[1]{\protect\rule{.1in}{.1in}}
\begin{document}
\title{Tunable spinful matter wave valve}
\author{Yan-Jun Zhao }
\affiliation{Beijing National Laboratory for Condensed Matter Physics, Institute of
Physics, Chinese Academy of Sciences, Beijing 100190, China}
\affiliation{School of Physical Sciences, University of Chinese Academy of Sciences,
Beijing 100190, China}
\author{Dongyang Yu }
\affiliation{Beijing National Laboratory for Condensed Matter Physics, Institute of
Physics, Chinese Academy of Sciences, Beijing 100190, China}
\affiliation{School of Physical Sciences, University of Chinese Academy of Sciences,
Beijing 100190, China}
\author{Lin Zhuang }
\affiliation{School of Physics, Sun Yat-Sen University, Guangzhou 510275, China}
\author{Wu-Ming Liu}
\email{wliu@iphy.ac.cn}
\affiliation{Beijing National Laboratory for Condensed Matter Physics, Institute of
Physics, Chinese Academy of Sciences, Beijing 100190, China}
\affiliation{School of Physical Sciences, University of Chinese Academy of Sciences,
Beijing 100190, China}

\begin{abstract}
We investigate the transport problem that a spinful matter wave is incident on
a strong localized spin-orbit-coupled Bose-Einstein condensate in optical
lattices, where the localization is admitted by atom interaction only existing
at one particular site, and the spin-orbit coupling arouse spatial rotation of
the spin texture. We find that tuning the spin orientation of the localized
Bose-Einstein condensate can lead to spin-nonreciprocal / spin-reciprocal
transport, meaning the transport properties are dependent on / independent of
the spin orientation of incident waves. In the former case, we obtain the
conditions to achieve transparency, beam-splitting, and blockade of the
incident wave with a given spin orientation, and furthermore the ones to
perfectly isolate incident waves of different spin orientation, while in the
latter, we obtain the condition to maximize the conversion of different spin
states. The result may be useful to develop a novel spinful matter wave valve
that integrates spin switcher, beam-splitter, isolator, and converter. The
method can also be applied to other real systems, e.g., realizing perfect
isolation of spin states in magnetism, which is otherwise rather difficult.

\end{abstract}
\revised{\today}

\startpage{1}
\endpage{ }
\maketitle

Ultracold atoms, where atom interaction and spin-orbit coupling (SOC) can be
artificially synthesized, are an ideal platform for simulating many-body
physics~\cite{Bloch2012NP,Jaksch2005AP,Goldman2016NP,Bloch2008RMP}. The
wave-particle duality points out that particles can behave like waves and also
vice verse~\cite{Cohen1977Book}. Thus, it is of interest to investigate the
matter wave properties of multiple cold atoms. Tunable via
magnetic~\cite{Donley2002Nature,Loftus2002PRL,Tiesinga1993PRA,Inouye1998Nature}
or optical~\cite{Fedichev1996PRL,Theis2004PRL} Feshbach resonance, the atom
interaction accounts for versatile intriguing phenomena featuring the
transport of\ spinless matter
waves~\cite{LiuWM2000PRL,Liang2005PRL,Morsch2006RMP,Miroshnichenko2010RMP,Kartashov2011RMP,Chien2015nphys,Poulsen2003PRA,Smerzi2003PRA,Vicencio2007PRL,Zhang2008EPJD,Arevalo2009PLA,Hennig2010PRA,Bai2015AnnPhys,Bai2016PRE}%
. Typically, a nonlinear impurity can blockade the transmission of a
perturbative incident wave~\cite{Vicencio2007PRL}. Besides, the discrete
breather, resulted from nonlinear lattices, can be partially transmitted, and
shifted by a moving breather~\cite{Hennig2010PRA}. Furthermore, when
asymmetric defects are immersed in the nonlinear lattices, the discrete
breather will be tilted, capably inducing the unidirectional transport of wave
packets~\cite{Bai2016PRE}. In spinor Bose-Einstein condensate (BEC), however,
the spin-dependent interaction can induce the non-Abelian Josephson
effect~\cite{Qi2009PRL}.

Meanwhile, as a key ingredient for spin Hall
effect~\cite{Kato2004Science,Konig2007Science} and topological
insulator~\cite{Kane2005PRL,Bernevig2006Science,Hsieh2008Nature}, SOC can be
generated through non-Abelian gauge fields induced by the space variation of
light~\cite{Dalibard2011RMP,Galitski2013Nature,Goldman2014RPP,Zhai2015RPP}. In
combination with atom interactions, SOC can affect the properties of localized
modes or solitons in cold atom BEC
\cite{Sakaguchi2014PRE1,Lobanov2014PRL,Sakaguchi2014PRE2,Belicev2015JPB,Gligoria2016PRB}%
. For example, Rashba SOC and cubic attractive interactions together can give
rise to two types of solitary-vortex complexes, respectively termed
semivortices and mixed modes~\cite{Sakaguchi2014PRE1}. Using the parity and
time reversal symmetries of a two-dimensional SOC BEC, localized solutions of
various families, including multipole and half-vortex solitons, can be
found~\cite{Lobanov2014PRL}. Compact localized states and discrete solitons
can coexist for nonlinear spinful waves on a flat-band network with
SOC~\cite{Gligoria2016PRB}. Although it has been
reported~\cite{Vicencio2007PRL} that the localized BEC can blockade the
propagation of an spinless incident wave, how to manipulate the transport of
spinful matter waves via tunable nonlinearity in SOC BEC\ in optical
lattices~\cite{Cole2012PRL,Xu2014PRA,Osterloh2005PRL,Hamner2015PRL,Jordens2008Nature,Goldman2009PRL,Goldman2009PRA,Radic2012PRL,Cai2012PRA}
remains an open problem.

In this Letter, we investigate the transport problem that a weak transmission
matter wave encounters a localized SOC BEC in optical lattices. In the
presence of SOC, both the transmission and localized modes exhibit
spin-rotation effect in the lattice space. The spin orientation, interaction
and atom number of the BEC can be artificially manipulated, which induces
tunable transport properties for incident waves with a definite spin
orientation. In general, if the BEC orients parallel to the incident waves, it
can behave like a spin switcher, beam-splitter, or isolator, while if they
orient perpendicular, the BEC behaves like a spin converter.

\begin{figure}[ptb]
\includegraphics[width=0.36\textwidth, clip]{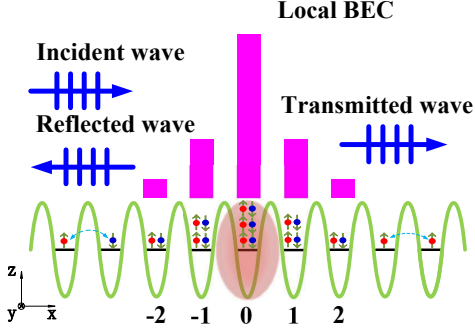}\caption{(color online)
Scattering process of the weak atomic matter wave incident on the strong BEC
localized in the vicinity of origin in optical lattices. Atoms are represented
by red and blue balls with internal spins shown by arrows. The strong
localized mode, whose magnitude is shown by magenta bars, is induced by
localized interactions around origin (attractive and idealized as a $\delta
$-type impurity). The spin-flipping hopping between adjacent sites is aroused
by SOC. The incident, reflected, and transmitted atoms with internal spins are
represented as plane waves.}%
\label{fig:model}%
\end{figure}

We consider the scattering process of the weak atomic matter wave incident on
a spin-orbit coupled localized BEC\ in optical lattices (see
Fig.~\ref{fig:model}). To create SOC, we can illuminate the $^{\text{87}}$Rb
bosonic particles by two counterpropagating Raman lasers with proper magnetic
bias, where the two internal atomic pseudo-spin-states are selected from
within the $^{\text{87}}$Rb $5S_{1/2}$, $F=1$ ground electronic manifold:
$\left\vert \uparrow\right\rangle =\left\vert F=1,m_{F}=0\right\rangle $
(pseudo-spin-up) and $\left\vert \downarrow\right\rangle =\left\vert
F=1,m_{F}=-1\right\rangle $ (pseudo-spin-down)~\cite{Lin2011Nature}. Besides,
the optical lattices can be generated through a standing wave in the large
detuning regime~\cite{Morsch2006RMP}. Moreover, the localization can be
induced by atom interactions concentrated on the vicinity of lattice
origin,\ which can be obtained by generating inhomogeneous s-wave scattering
length of atoms via tuning
magnetic~\cite{Donley2002Nature,Loftus2002PRL,Tiesinga1993PRA,Inouye1998Nature}
or optical~\cite{Fedichev1996PRL,Theis2004PRL} Feshbach
resonance.

In a mean-field form, the system can be well described by the Hamiltonian
\begin{equation}
H\!\!=\!\!-J\!\sum_{n}(\psi_{n+1}^{\dag}\!R\psi_{n}\!+\!\text{H.c.)}-\frac
{J}{2}\!\!\sum_{n\sigma\sigma^{\prime}}\!\delta_{n0}\gamma_{\sigma
\sigma^{\prime}}\!\left\vert \psi_{n\sigma}\!\right\vert ^{2}\!\left\vert
\psi_{n\sigma^{\prime}}\!\right\vert ^{2}\!. \label{eq:system_Hamiltonian}%
\end{equation}
where $\psi_{n}=(\psi_{n\uparrow},\psi_{n\downarrow})^{T}$ represents the
macroscopic wave function of the BEC. The lattice potential well is deep
enough to only involve the hopping between nearest neighbours. Concretely, the
spin-conserving (spin-flipping) hopping is characterized by the diagonal
(off-diagonal) terms of the spin-rotation operator $R=\exp\left(  -i\sigma
_{y}\alpha\right)  $~\cite{Cole2012PRL,Xu2014PRA} which arises from the
non-Abelian potential $\mathbf{A}\!=\!\left(  \alpha\sigma_{y},0,0\right)  $
through Peierls substitution~\cite{Hofstadter1976PRB}. The parameter $\alpha$
is a ratio: $\alpha=\pi k_{\text{soc}}/k_{\text{ol}}$, where $k_{\text{soc}}$
describes the momentum transfer from the Raman lasers and $k_{\text{ol}}$ is
the wave vector of the optical lattice~\cite{Radic2012PRL,Cai2012PRA}. Not
losing generality, we set the hopping strength $J=1$ hereafter.

The localized interactions is idealized as a $\delta$-type impurity, which
vanishes except at $n=0$. We choose the intraspecies interaction
$\gamma_{\uparrow\uparrow}=\gamma_{\downarrow\downarrow}=\gamma$ and the
interspecies interaction $\gamma_{\uparrow\downarrow}=\gamma_{\downarrow
\uparrow}=\lambda\gamma$~\cite{Cole2012PRL,Xu2014PRA} with $\gamma,\lambda>0$
(attractive interaction). Besides, $\gamma,\lambda\gamma\ll1$ is hypothesized
to validate the mean-field approach. Hereafter, $\gamma$ and $\lambda$ are
called interaction strength and interaction ratio, respectively.

\begin{figure}[ptb]
\includegraphics[width=0.21\textwidth, clip]{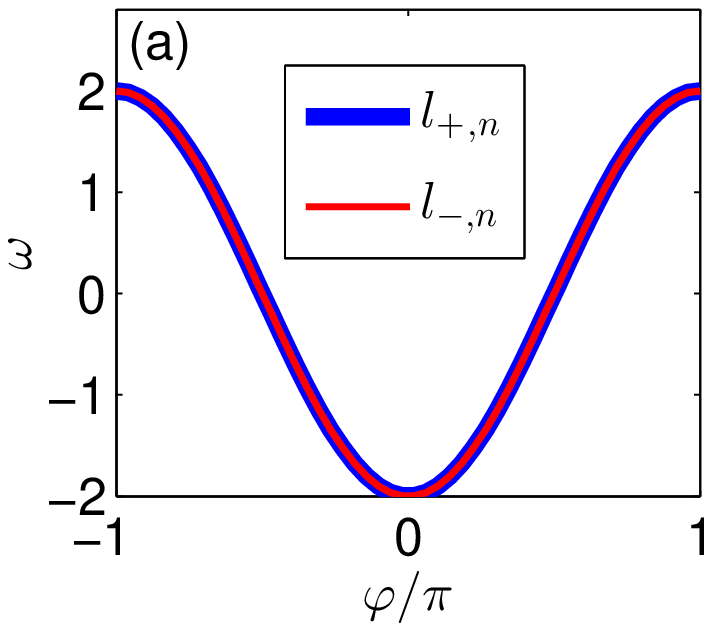}\includegraphics[width=0.21\textwidth, clip]{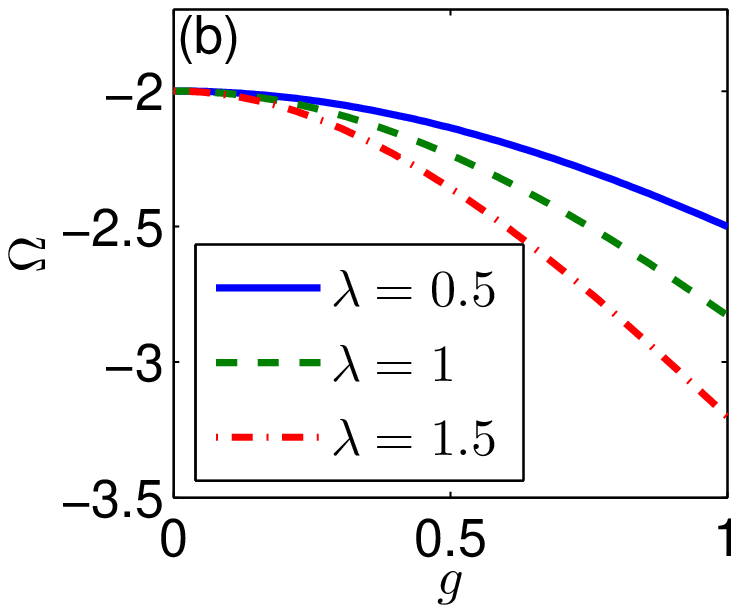}\newline%
\includegraphics[width=0.42\textwidth, clip]{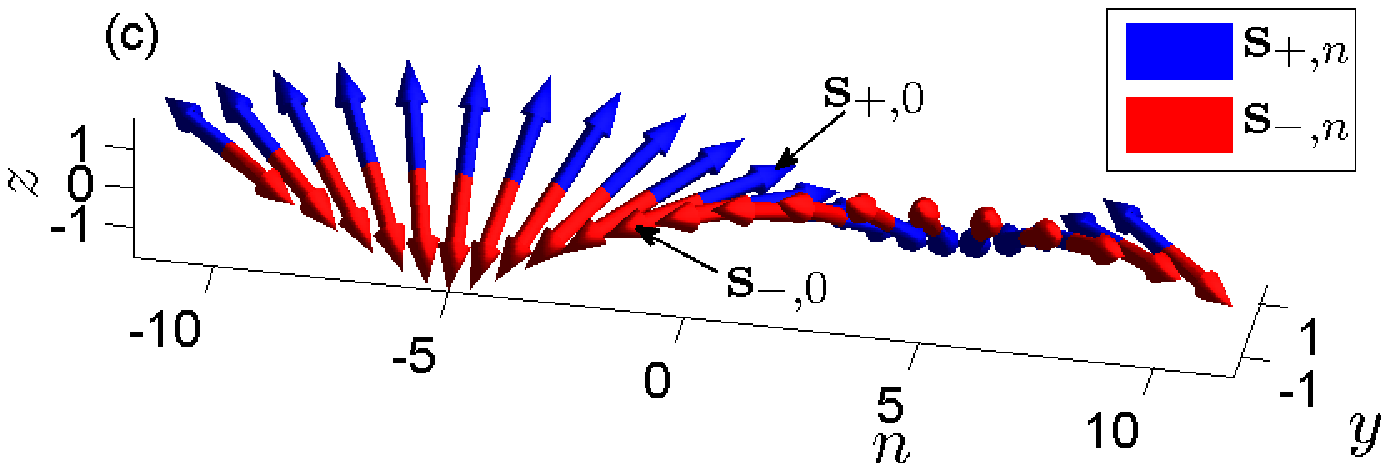}\newline%
\includegraphics[width=0.42\textwidth, clip]{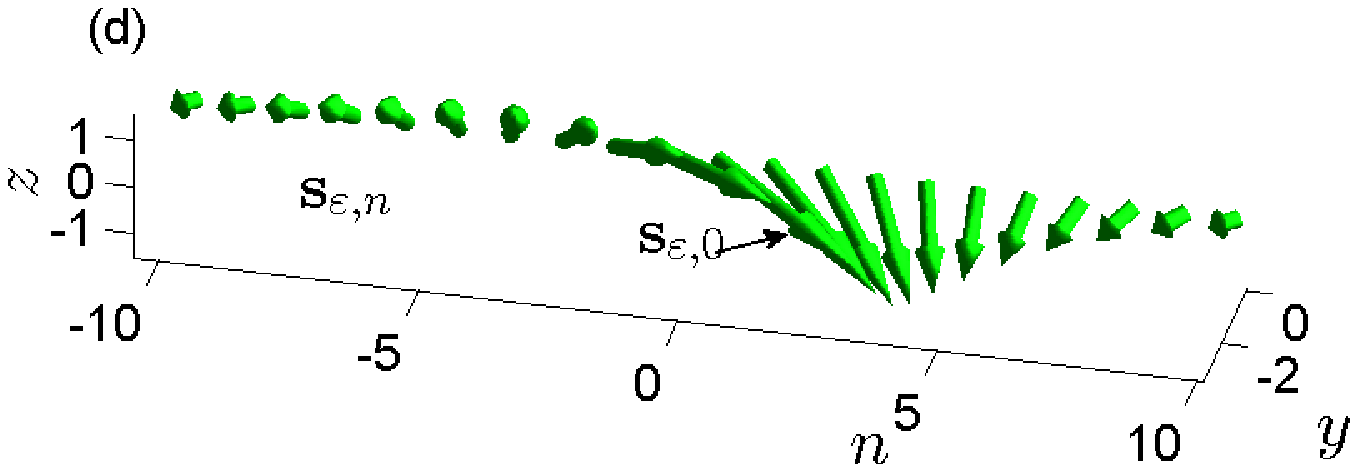}\caption{(color online)
(a) Dispersion relation of the transmission modes $l_{+,n}$ (solid blue) and
$l_{-,n}$ (solid red). (b) Eigenenergy $\Omega$ of the strong localized mode
against $g$ for $\lambda$ taking $0.5$ (solid blue), $1$ (dashed green), and
$1.5$ (dash-dotted red), respectively. (c) Spin texture of the transmission
states: $\mathbf{s}_{+,n}=l_{+,n}^{\dag}\boldsymbol{\sigma}l_{+,n}$ (blue) and
$\mathbf{s}_{-,n}=l_{-,n}^{\dag}\boldsymbol{\sigma}l_{-,n}$ (red). We can see
spin rotating with $y$-axis when $n$ changes, where $\alpha$ is assigned
$\pi/20$, resulting in a rotation period $\pi/\alpha=20$. Besides, $a=\pi/4$
and $b=\pi/2$, thus $\mathbf{s}_{\pm,0}$ directing $\pm\left(  1,1,0\right)
$. (d) Spin texture of the strong localized state: $\mathbf{s}_{\varepsilon
,n}=d_{n}^{\dag}\boldsymbol{\sigma}d_{n}$, where $\Omega=-2.01$, $g=0.9$ and
$\gamma=0.002$. We specify $\varepsilon=$ $\pi/4$, thus $\mathbf{s}%
_{\varepsilon,0}$ directing $(1,-1,0)$. The rotating of $\mathbf{s}%
_{\varepsilon,n}$ is similar to $\mathbf{s}_{\pm,n}$. Spin textures for
$\alpha=\pi/10$ are also plotted for comparision~\cite{SM}.}%
\label{fig:stronglocalmode_iii}%
\end{figure}

We now seek the transmission modes using Gross-Pitaevskii equation
$i\partial\psi_{n}/\partial t\!=\!\partial H/\partial\psi_{n}^{\ast}$. The
transmission mode $l_{n}\exp\left(  -i\omega t\right)  $ is dominated by the
free Hamiltonian of atoms [first term of Eq.~(\ref{eq:system_Hamiltonian})].
Thus, setting $U=0$, we obtain the stationary solution $l_{\pm,n}%
\!=\!e^{in\varphi}R^{n}l_{\pm}$ with the dispersion relation $\omega
\!=\!-2\cos\varphi$ [see Fig.~\ref{fig:stronglocalmode_iii}(a)]. The energy
$\omega$ is irrelevant with the spin freedom $l_{\pm}$, which can be generally
given by $l_{+}=\cos\left(  a/2\right)  u_{+}+e^{ib}\sin\left(  a/2\right)
u_{-}$ and $l_{-}\!\!=\!-e^{-ib}\sin\left(  a/2\right)  u_{+}+\cos\left(
a/2\right)  u_{-}$, where $b\in\left[  0,\pi\right]  $, and $u_{\pm}=\left(
1,\pm i\right)  ^{\mathrm{T}}$ (eigenstates of $\sigma_{y}$). The spin
orientation of $l_{\pm,n}$ is $\mathbf{s}_{\pm,n}=l_{\pm,n}^{\dag
}\boldsymbol{\sigma}l_{\pm,n}=\pm2[\sin a\sin\left(  b+2n\alpha\right)
\mathbf{e}_{x}+\cos a\mathbf{e}_{y}+\sin a\cos\left(  b+2n\alpha\right)
\mathbf{e}_{z}]$ [see Fig.~\ref{fig:stronglocalmode_iii}(c)]. Given definite
energy $\omega$, four degenerate transmission states are resulted in:
\begin{align}
l_{n}^{\left(  1\right)  } &  =e^{in\varphi}R^{n}l_{+},l_{n}^{\left(
2\right)  }=e^{-in\varphi}R^{n}l_{+},\nonumber\\
l_{n}^{\left(  3\right)  } &  =e^{in\varphi}R^{n}l_{-},l_{n}^{\left(
4\right)  }=e^{-in\varphi}R^{n}l_{-},
\end{align}
where $\varphi\!=\!\arccos\left(  -\omega/2\right)  \!\in\![0,\pi]$ is
explicitly hypothesized.

By contrast, the localized mode $d_{n}\exp\left(  -i\Omega t\right)  $ is
dominated by the full Hamiltonian $H$. Thus, we obtain the stationary solution
$d_{n}\!=\!\sqrt{g/\gamma}\kappa^{\left\vert n\right\vert }R^{n}\mathcal{E}$
and eigenenergy $\Omega=-\sqrt{(1+\lambda)^{2}g^{2}+4}$~[see
Fig.~\ref{fig:stronglocalmode_iii}(b)]. Here, $g$ is called localization
grade, which determines the spatial decay rate $\kappa\!=\!(-\Omega
-\sqrt{\Omega^{2}-4})/2$. Besides, the spin part $\mathcal{E}=\left(
e^{i\varepsilon},1\right)  ^{T}$ determines the spin texture of $d_{n}$, i.e.,
$\mathbf{s}_{\varepsilon,n}=d_{n}^{\dag}\boldsymbol{\sigma}d_{n}%
\mathcal{=}2\left(  g/\gamma\right)  \kappa^{2\left\vert n\right\vert }%
[\cos\varepsilon\cos\left(  2n\alpha\right)  \mathbf{e}_{x}-\sin
\varepsilon\mathbf{e}_{y}-\cos\varepsilon\sin\left(  2n\alpha\right)
\mathbf{e}_{z}]$~[see Fig.~\ref{fig:stronglocalmode_iii}(d)]. Due to the
effect of SOC, both $\mathbf{s}_{\pm,n}$ and $\mathbf{s}_{\varepsilon,n}$
rotates with $y$-axis as $n$ changes, where the winding number per increment
of lattice site is $\alpha/\pi$. The atom number is $N_{\text{at}}%
=-2\Omega/\left(  1+\lambda\right)  \gamma$, meaning implicitly that $g$ is
tunable via modifying the interaction strength $\gamma$, interaction ratio
$\lambda$, or atom number $N_{\text{at}}$.

We now derive the spinful-wave-strong-localized-BEC-interaction via
substituting $\psi_{n}=\phi_{n}+\Psi_{n}$ into the dynamical equation
$i\partial\psi_{n}/\partial t=\partial H/\partial\psi_{n}^{\ast}$. Here,
$\Psi_{n}=d_{n}e^{-i\Omega t}$ is the strong localized BEC while $\phi_{n}$ is
weak and represents incident and stimulated waves. Rigorously, we assume
$\left\vert \phi_{0\sigma}\right\vert \ll\left\vert \Phi_{0\sigma^{\prime}%
}\right\vert =\sqrt{g/\gamma}$, thus resulting in the linearized dynamical
equation with respect to $\phi_{n}$:
\begin{equation}
i\frac{\partial\phi_{n}}{\partial\tau}=-R\phi_{n-1}-R^{\dag}\phi_{n+1}%
-\delta_{n0}(R_{-}\phi_{0}+R_{\tau}R_{+}\phi_{0}^{\ast}).
\label{eq:Lin_dynamics}%
\end{equation}
The parameters $R_{\pm}=g[\lambda+2+\lambda\left(  \cos\varepsilon\pm
\sigma_{y}\sin\varepsilon\right)  ]$ and $R_{\tau}=\exp\left[  i\left(
\varepsilon-2\Omega\tau+\sigma_{z}\varepsilon\right)  \right]  $ quantify the
non-Abelian potential generated by the strong localized BEC at origin. Once
encountered, the potential will scatter off a spinful wave or flip its spin
which will otherwise propagate freely governed by the first two terms in
Eq.~(\ref{eq:Lin_dynamics}).

Motivated by the presence of $R_{\tau}$, we treat Eq.~(\ref{eq:Lin_dynamics})
with the ansatz $\phi_{n}=p_{n}e^{-i\omega\tau}+q_{n}e^{-i\nu\tau}$ where
$\nu=2\Omega-\omega$. The fact $\omega\in\left[  -2,2\right]  $ determines
$p_{n}$'s nature of being extended states. But $q_{n}$ must be localized since
$\nu<-2$ is caused. This treatment results in the coupled equations that
feature the interplay between both states:
\begin{align}
\omega p_{n}  &  =-Rp_{n-1}-R^{\dag}p_{n+1}-\delta_{n0}(R_{-}p_{0}%
+R_{\varepsilon}R_{+}q_{0}^{\ast}),\nonumber\\
\nu q_{n}  &  =-Rq_{n-1}-R^{\dag}q_{n+1}-\delta_{n0}(R_{-}q_{0}+R_{\varepsilon
}R_{+}p_{0}^{\ast}), \label{eq:qn}%
\end{align}
where $R_{\varepsilon}=\exp\left[  i\left(  \varepsilon+\sigma_{z}%
\varepsilon\right)  \right]  $.

Nevertheless, defining a transport process needs the incident wave specified
in detail. Note $\varphi\in\lbrack0,\pi]$ makes the group velocity $v_{j}$ of
transmission mode $l_{n}^{\left(  j\right)  }$ meet $v_{1,3}=2\sin\varphi
\geq0$ and $v_{2,4}=-2\sin\varphi\leq0$. Thus, we represent the
negative-incident waves as $L_{n}^{\left(  j\right)  }=l_{n}^{\left(
j\right)  }\theta\left(  -n-1\right)  $, $j=1,3$ [$\theta\left(  \cdot\right)
$ is the Heaviside step function], but the positive-incident ones as
$L_{n}^{\left(  j\right)  }=l_{n}^{\left(  j\right)  }\theta\left(  n\right)
$, $j=2,4$. The solution of $p_{n}$ and $q_{n}$ can accordingly take the forms
$p_{n}^{\left(  j\right)  }\left(  \alpha\right)  =L_{n}^{(j)}+(S_{2j}%
l_{n}^{\left(  2\right)  }+S_{4j}l_{n}^{\left(  4\right)  })\theta
(-n-1)+(S_{1j}l_{n}^{\left(  1\right)  }+S_{3j}l_{n}^{\left(  3\right)
})\theta(n)$ and $q_{n}^{\left(  j\right)  }\left(  \alpha\right)  =R^{n}%
q_{0}^{\left(  j\right)  }\chi^{\left\vert n\right\vert }$. However, the
isotropy of the transport process (e.g., $S_{12}=S_{21}$) can be
justified~\cite{SM}. Hence, only the cases of $j=1,3$ merit investigation. By
inserting $p_{n}=p_{n}^{\left(  j\right)  }\left(  \alpha\right)  $ and
$q_{n}=q_{n}^{\left(  j\right)  }\left(  \alpha\right)  $ into
Eq.~(\ref{eq:qn}) for $n$ taking $-1$, $0$, and $1$, respectively, we obtain
the scattering coefficients $S_{j^{\prime}j}$ for $j=1,3$:
\begin{align}
S_{11}  &  =S_{21}+1=\frac{i\tilde{\varphi}\left(  i\tilde{\varphi}%
+X+YC_{Y}\right)  }{\left(  i\tilde{\varphi}+X\right)  ^{2}-Y^{2}},\nonumber\\
S_{31}  &  =S_{41}=\frac{i\tilde{\varphi}\left(  iY\right)  \left(
ie^{ib}\sin a\sin\varepsilon-C_{\varepsilon}\cos\varepsilon\right)  }{\left(
i\tilde{\varphi}+X\right)  ^{2}-Y^{2}},\nonumber\\
S_{33}  &  =S_{43}+1=\frac{i\tilde{\varphi}\left(  i\tilde{\varphi}%
+X-YC_{Y}\right)  }{\left(  i\tilde{\varphi}+X\right)  ^{2}-Y^{2}},\nonumber\\
S_{13}  &  =S_{23}=\frac{i\tilde{\varphi}\left(  iY\right)  \left(
ie^{-ib}\sin a\sin\varepsilon+C_{\varepsilon}^{\ast}\cos\varepsilon\right)
}{\left(  i\tilde{\varphi}+X\right)  ^{2}-Y^{2}}, \label{eq:S_13}%
\end{align}
Here, $\tilde{\varphi}=2g^{-1}\sin\varphi$, $C_{Y}=\sin\varepsilon\cos
a-\cos\varepsilon\sin a\sin b$, $C_{\varepsilon}=\cos^{2}\left(  a/2\right)
+e^{i2b}\sin^{2}\left(  a/2\right)  $, $X\equiv X(\omega)$, and $Y\equiv
Y(\omega)$~\cite{SM}. As $S_{2j}$ $(S_{4j})$ can be deduced from $S_{1j}$
$(S_{3j})$, we hereafter only discuss $S_{j^{\prime}j}$ for $j^{\prime}%
,j\in\{1,3\}$.

Since $\mathbf{s}_{\pm,n}$ represent the spin orientation of $l_{n}^{\left(
1\right)  }$ and $l_{n}^{\left(  3\right)  }$, one notes spin-nonreciprocal
transport ($\left\vert S_{11}\right\vert \neq\left\vert S_{33}\right\vert $)
can be achieved when $C_{Y}\neq0$ (or rather, $\tan\varepsilon\neq\tan a\sin
b)$, although $\left\vert S_{31}\right\vert =\left\vert S_{13}\right\vert $
always holds~\cite{SM}. Furthermore, we can justify the transparency
($S_{jj}=1$) and blockade ($S_{jj}=0$) are realizable only when $C_{Y}=\mp1$
(i.e., $b=\pi/2$ and $\varepsilon=a\mp\pi/2$), meaning $\mathbf{s}_{\pm,0}$
orient within the $xoy$ plane, and $\mathbf{s}_{\varepsilon,n}$
orients~identical to $\mathbf{s}_{+,n}$ or $\mathbf{s}_{-,n}$. If
$\mathbf{s}_{\varepsilon,n}$ orients identical to $\mathbf{s}_{+,n}$
($\mathbf{s}_{-,n}$), the incident wave $L_{n}^{\left(  1\right)  }$
[$L_{n}^{\left(  3\right)  }$] and $L_{n}^{\left(  3\right)  }$ [$L_{n}%
^{\left(  1\right)  }$] will undergo transparency at T1 and T2, and blockade
at B1 and B2, where
\begin{align}
\text{T1}  &  \text{: }\mu=\frac{3}{2}\left(  \lambda+1\right)  ,\text{B1:
}\mu=2\lambda+2,\nonumber\\
\text{T2}  &  \text{: }\mu=-\frac{1}{2}\left(  \lambda-3\right)  \left(
\lambda+1\right)  ,\text{B2: }\mu=2,
\end{align}
and $\mu=\sqrt{(2\Omega-\omega)^{2}-4}/g$. When the energy $\omega$ deviates
from T1 (T2) and B1 (B2), $L_{n}^{\left(  1\right)  }$ [$L_{n}^{\left(
3\right)  }$] will undergo partial transmission, which can be interpreted as
the beam-splitting effect. Moreover, $C_{Y}=\mp1$ causes $S_{31}=S_{13}=0$,
signifying no conversion between $l_{n}^{\left(  1\right)  }$ and
$l_{n}^{\left(  3\right)  }$ in the output fields.

\begin{figure}[ptbh]
\includegraphics[width=0.166\textwidth, clip]{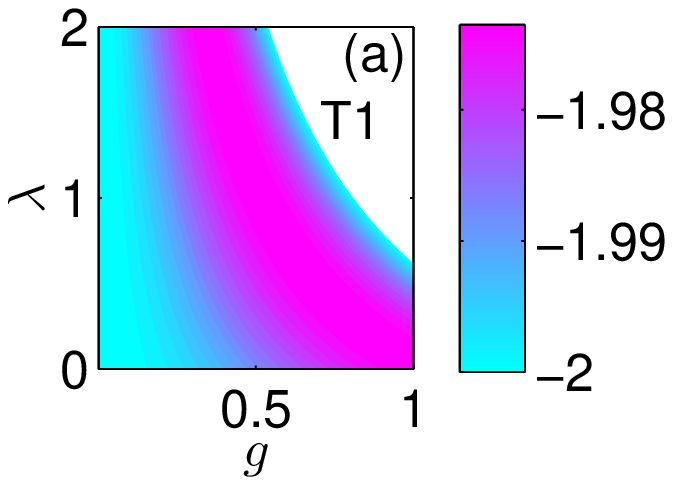}\includegraphics[width=0.166\textwidth, clip]{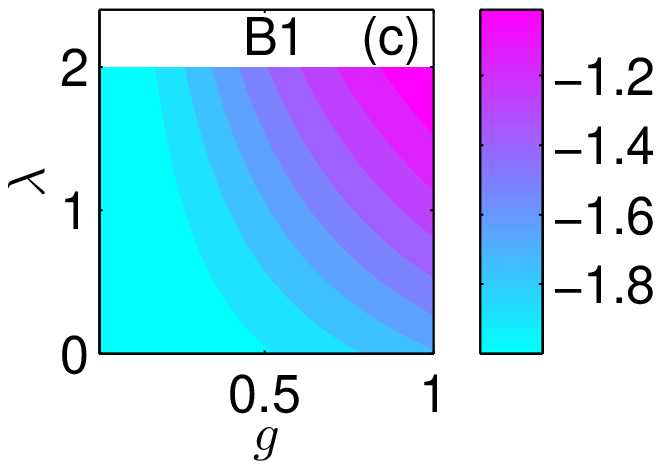}\includegraphics[width=0.166\textwidth, clip]{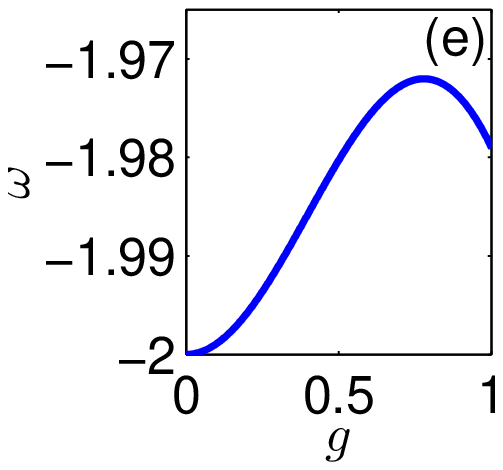}\newline%
\includegraphics[width=0.166\textwidth,clip]{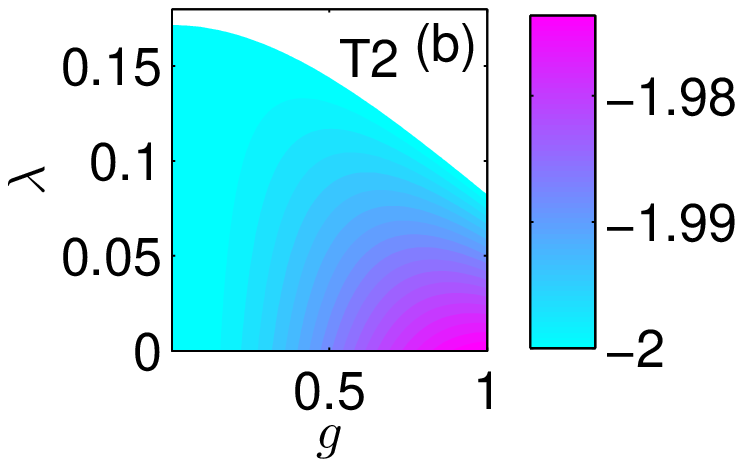}\includegraphics[width=0.166\textwidth, clip]{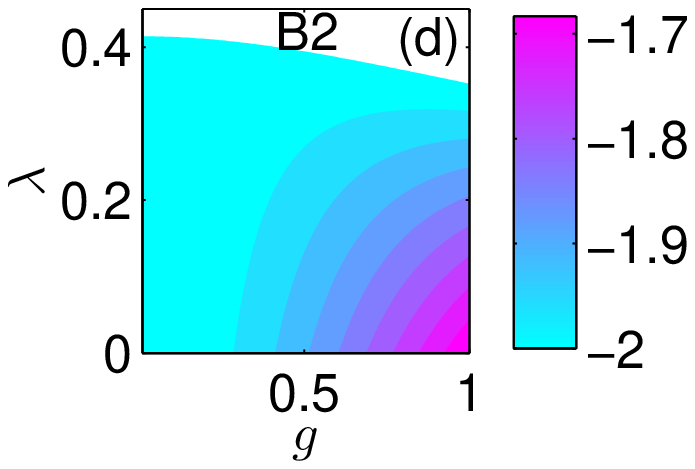}\includegraphics[width=0.166\textwidth, clip]{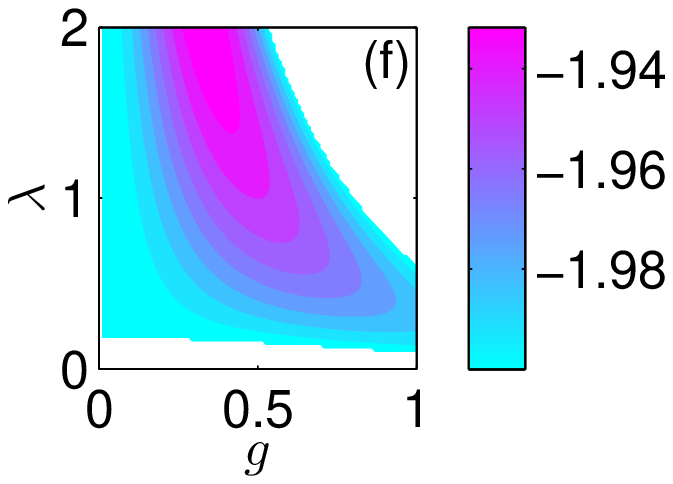}\caption{(color
online) Energy $\omega$ against interaction ratio $\lambda$ and localization
grade $g$ at transparency points (a) T1, (b) T2, and blockade points (c) B1,
(d) B2. (e) Energy $\omega$ against $g$ at the isolation point. (f) Energy
$\omega$ plotted against $g$ and $\lambda$ at the maximum spin conversion
point.}%
\label{fig:TB}%
\end{figure}

Until now, we identify non-reciprocal transport behaviours depending on
different spin orientation of the incident wave. In Figs.~\ref{fig:TB}%
(a)-\ref{fig:TB}(d), we show the controllability of transparency and blockade
points. In Figs.~\ref{fig:Sim}(a)-\ref{fig:Sim}(f), we present the simulation
result using exact dynamical equation $i\partial\psi_{n}/\partial t=\partial
H/\partial\psi_{n}^{\ast}$, where the perturbative part is initialized with a
Gaussian profile: $\phi_{n}\left(  0\right)  =s_{0}\exp[-s_{p}\left(
n-n_{0}\right)  ^{2}]L_{n}^{\left(  j\right)  }$. We also specify $b=\pi/2$
and $\varepsilon=a-\pi/2$ such that $\mathbf{s}_{\mathcal{\varepsilon}}$ is
identical (opposite) to $\mathbf{s}_{\mathcal{+},n}$ ($\mathbf{s}%
_{\mathcal{-},n}$). Through the simulation results, tunable transport is shown
from transparency [Figs.~\ref{fig:Sim}(a) and \ref{fig:Sim}(b)], beam
splitting [Figs.~\ref{fig:Sim}(c) and \ref{fig:Sim}(d)], to blockade
[Figs.~\ref{fig:Sim}(e) and \ref{fig:Sim}(f)].%

\begingroup\begin{figure*}[ptb]
\includegraphics[width=0.22\textwidth, clip]{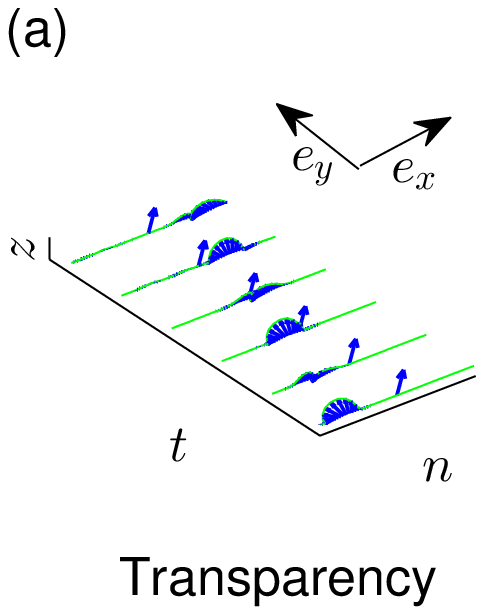}
\includegraphics[width=0.22\textwidth, clip]{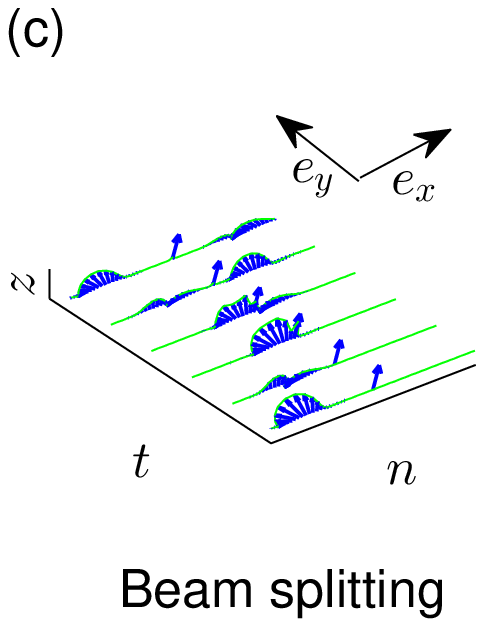}
\includegraphics[width=0.22\textwidth, clip]{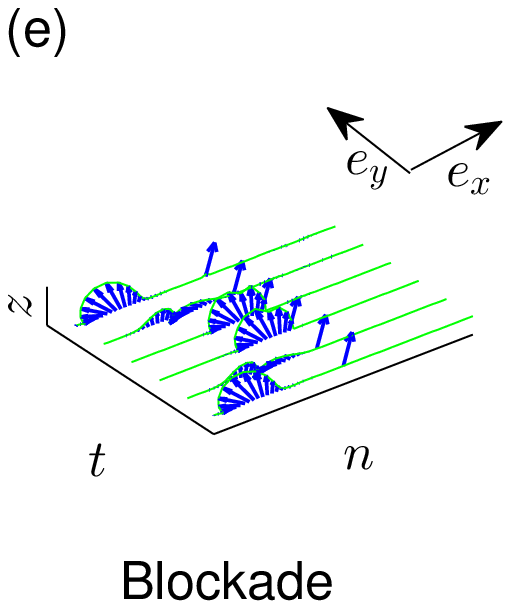}
\includegraphics[width=0.22\textwidth, clip]{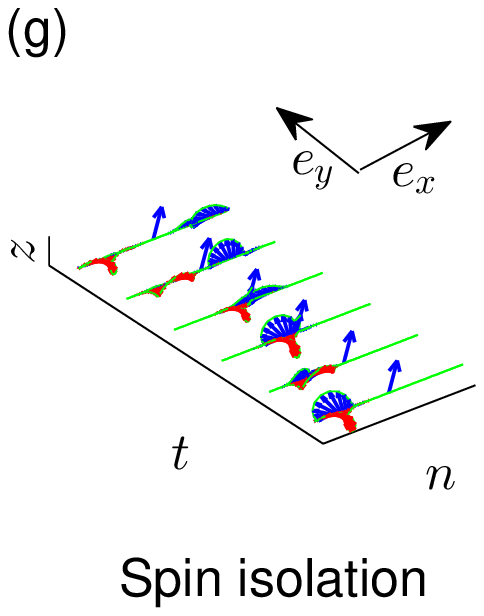}\\
\includegraphics[width=0.22\textwidth, clip]{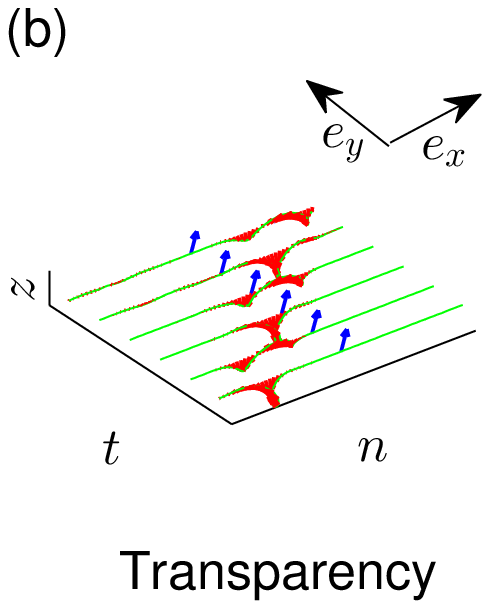}
\includegraphics[width=0.22\textwidth, clip]{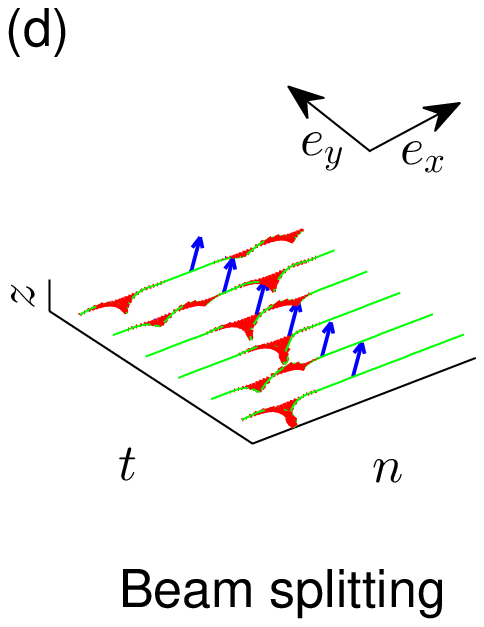}
\includegraphics[width=0.22\textwidth, clip]{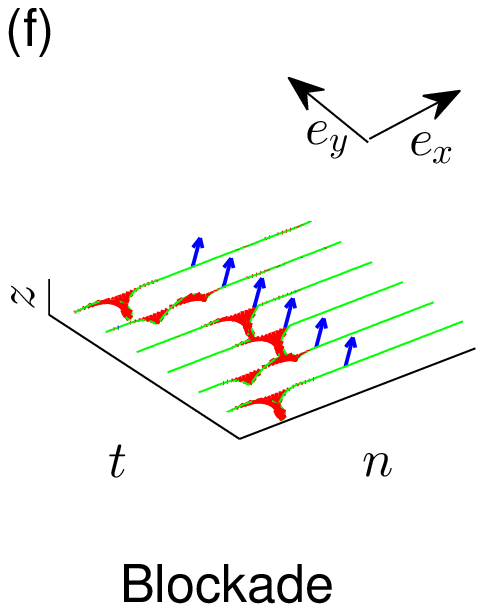}
\includegraphics[width=0.22\textwidth, clip]{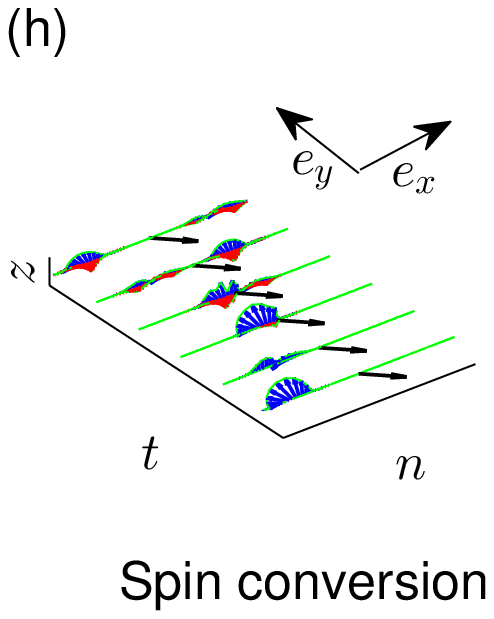}
\caption
{(color online) Simulated time evolution of the transport process. (a) and (b) Transparency with localization grade
$g=0.9$ and interaction ratio $\lambda
=0.025$. (c) and (d) Beam splitting with $g=0.69$ and
$\lambda=0.1$. (e) and (f) Blockade with $g=0.75$ and $\lambda=0.1$. (g) Spin
isolation with $g=0.7788$ and $\lambda=1/3$. (h) Spin conversion with
$g=0.5$ and $\lambda=1$. The blue (red) arrows mean spin orientation along
$\mathbf{s}_{+,n}$ ($\mathbf{s}_{-,n}%
$), while the black ones mean spin orientation perpendicular to $\mathbf
{s}_{\pm,n}$.}
\label{fig:Sim}
\end{figure*}\endgroup

Now we explore the possibility of achieving perfect isolation of different
spin states, that is, making one spin state fully transmitted and the other
totally reflected. To this end, there are two possible situations. (i) T1 and
B2 overlaps, yielding $\lambda=1/3$; (ii) T2 and B1 overlaps, yielding
$\lambda=-1$. The case $\lambda=-1$ exceeds the scope of the present
discussion. Therefore, we only concentrate on $\lambda=1/3$, with energy
$\omega$ determined by $\mu=2$. In this case, the incident wave $L_{n}%
^{\left(  j\right)  }$ will be fully transmitted if orienting identically to
that of $d_{n}$, otherwise be totally reflected. The controllability of the
isolation point is shown in Fig.~\ref{fig:TB}(e). The simulation result using
the exact dynamical equation is presented in Fig.~\ref{fig:Sim}(g).

We are also curious about the conversion between $l_{n}^{\left(  1\right)  }$
and $l_{n}^{\left(  3\right)  }$, which is measured by scattering coefficients
$S_{31}$ and $S_{13}$. In contrast to the transparency and blockade cases, the
strong localized mode should fulfill the condition $\tan\varepsilon=\tan a\sin
b$~\cite{SM}. It means $\mathbf{s}_{\varepsilon}$ must orient perpendicular to
$\mathbf{s}_{\pm,n}$. Meanwhile, the relation $S_{11}=S_{33}$ is caused,
implying spin-reciprocal transport behaviours which are independent of the
spin orientation of incident waves. Moreover, the energy $\omega$ of the
incident wave is required to satisfy $4-\omega^{2}=g^{2}\left(  Y^{2}%
-X^{2}\right)  $ [see Fig.~\ref{fig:TB}(f)]. In this case, maximum conversion
efficiency can be achieved as $\left\vert S_{31}\right\vert =\left\vert
S_{13}\right\vert =1/2$. The simulation result using the exact dynamical
equation is presented in Fig.~\ref{fig:Sim}(h).

In experiment, the incident $^{\text{87}}$Rb atoms can acquire the
quasimomentum $\varphi$ via phase imprinting method (i.e., using an
off-resonant light pulse to generate a proper light-shift potential which
dominates the evolution of the initial BEC
wavepacket)~\cite{Denschlag2000Science}, Bragg scattering, or simply
acceleration of the matter-wave probe in an external potential. The spin of
the BEC can be manipulated by Rabi oscillation induced by Raman laser pulses
that couple internal spin states with two-photon resonance. To measure the
scattering atoms, we first use a Stern--Gerlach gradient to separate atoms of
different spin states whose quantity can be further calculated via absorption
imaging~\cite{Lin2011Nature}.

In conclusion, we investigate the transport of a spinful matter wave scattered
by a strong localized BEC, in which the matter wave undergoes spin rotation as
lattice site changes due to the presence of SOC, and the strong localized BEC
generates an effective non-Abelian potential to the spinful wave which
furthermore impacts its transport behaviour. Tuning the BEC spin orientation
to orient parallel to that of the incident wave, we can achieve transparency,
blockade, and beam splitting of the incident wave. However, both the
transparency and blockade points are different for two incident waves with
opposite spin orientation. Thus, it is feasible to isolate two waves of
different spin orientation. In contrast, the maximum conversion between matter
waves with opposite spin orientation can also be achieved once the localized
BEC is tuned to orient perpendicular to the incident waves. The result may be
heuristic for developing a novel spinful matter wave valve that integrates
spin switcher, beam-splitter, isolator, and converter on a single atomic chip.
The proposal extends the atomtronics~\cite{Pepino2009PRL} to a spinful case,
i.e., a matter-wave version of spintronics, which is believed to give insights
in many quantum-based applications (e.g., gravitometry, magnetometry, etc).
Also, our proposal may facilitate the perfect isolation of spin states in
magnetism, which is otherwise rather difficult.

We are grateful to Ru-Quan Wang, Zai-Dong Li, Yi Zheng, Ji Li, Dong-Yang Jing,
Wen-Xiang Guo, Huan-Yu Wang, Wen-Xi Lai, Li Dai, and Chao-Fei Liu for helpful
discussions. This work is supported by the National Key R\&D Program of China
under grants Nos. 2016YFA0301500, NSFC under grants Nos. 11434015, 61227902,
11611530676, SPRPCAS under grants No. XDB01020300, XDB21030300, China
Postdoctoral Science Foundation under grant No. 2017M620945.

\bibliographystyle{apsrev}
\bibliography{SOT-references}

\end{document}


\title{Supplementary Material for \textquotedblleft Tunable spinful matter wave
valve\textquotedblright}
\author{Yan-Jun Zhao }
\affiliation{Beijing National Laboratory for Condensed Matter Physics, Institute of
Physics, Chinese Academy of Sciences, Beijing 100190, China}
\affiliation{School of Physical Sciences, University of Chinese Academy of Sciences,
Beijing 100190, China}
\author{Dongyang Yu }
\affiliation{Beijing National Laboratory for Condensed Matter Physics, Institute of
Physics, Chinese Academy of Sciences, Beijing 100190, China}
\affiliation{School of Physical Sciences, University of Chinese Academy of Sciences,
Beijing 100190, China}
\author{Lin Zhuang }
\affiliation{School of Physics, Sun Yat-Sen University, Guangzhou 510275, China}
\author{Wu-Ming Liu}
\email{wliu@iphy.ac.cn}
\affiliation{Beijing National Laboratory for Condensed Matter Physics, Institute of
Physics, Chinese Academy of Sciences, Beijing 100190, China}
\affiliation{School of Physical Sciences, University of Chinese Academy of Sciences,
Beijing 100190, China}
\keywords{cold atoms, blockade, amplification, spin-orbit coupling, nonlinear interaction}
\pacs{03.75.Lm, 05.45. a, 05.60.Gg, 42.25.Bs}
\revised{\today}

\startpage{1}
\endpage{ }
\maketitle

\section{Influence of the spin-orbit coupling}

\begin{figure}[h]
\includegraphics[width=0.35\textwidth, clip]{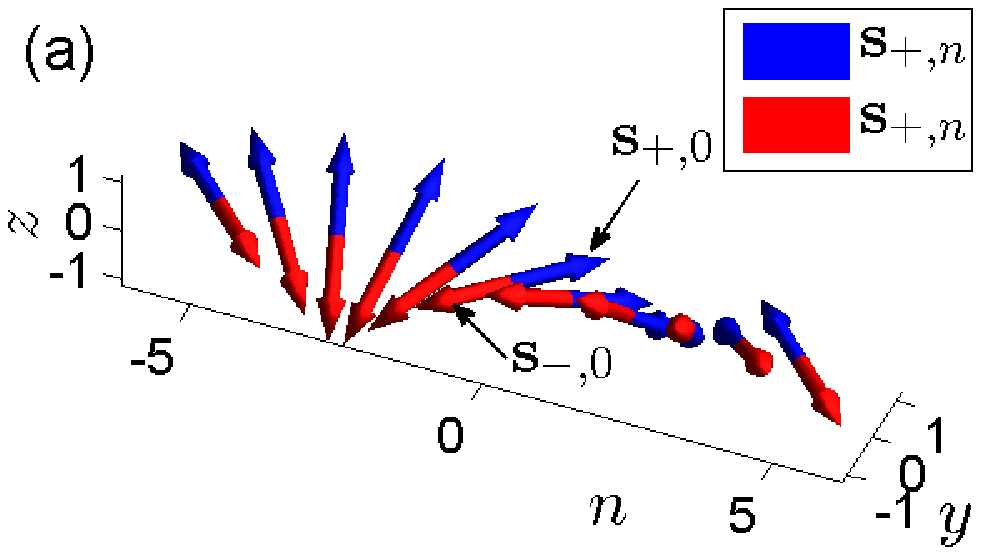}\includegraphics[width=0.35\textwidth, clip]{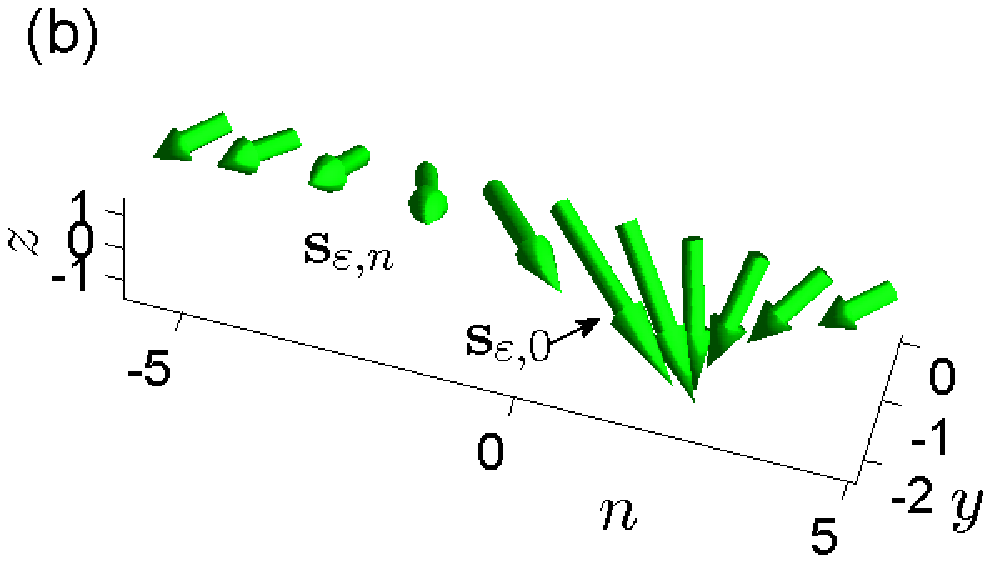}\caption{(color
online) (a) Spin texture of the transmission states: $\mathbf{s}_{+,n}%
=l_{+,n}^{\dag}\boldsymbol{\sigma}l_{+,n}$ (blue) and $\mathbf{s}%
_{-,n}=l_{-,n}^{\dag}\boldsymbol{\sigma}l_{-,n}$ (red). We can see spin
rotating with $y$-axis when $n$ changes, where $\alpha$ is assigned $\pi/10$,
resulting in a rotation period $\pi/\alpha=10$. Besides, $a=\pi/4$ and
$b=\pi/2$, thus $\mathbf{s}_{\pm,0}$ directing $\pm\left(  1,1,0\right)  $.
(b) Spin texture of the strong localized state: $\mathbf{s}_{\varepsilon
,n}=d_{n}^{\dag}\boldsymbol{\sigma}d_{n}$, where $\Omega=-2.01$, $g=0.9$ and
$\gamma=0.002$. We specify $\varepsilon=$ $\pi/4$, thus $\mathbf{s}%
_{\varepsilon,0}$ directing $(1,-1,0)$. The rotating of $\mathbf{s}%
_{\varepsilon,n}$ is similar to $\mathbf{s}_{\pm,n}$. }%
\label{fig:stronglocalmode_iii_suppl}%
\end{figure}

The spin-orbit coupling is characterized by the parameter $\alpha$, which, as
mentioned in the main text, determines the rotation period of the spin
texture: $\pi/\alpha$. In contrast to $\alpha=\pi/20$ in the main text, here
we show the case $\alpha=\pi/10$ in Fig.~\ref{fig:stronglocalmode_iii_suppl}.

\section{Justification of the isotropy of the transport process}

We now justify the transport process must be isotropic. Suppose the solution
is $p_{n}=p_{n}^{\left(  j\right)  }\left(  \alpha\right)  $ and $q_{n}%
=q_{n}^{\left(  j\right)  }\left(  \alpha\right)  $, then we can obtain the
coupled equations for the solution:%
\begin{align}
\omega p_{n}^{\left(  j\right)  }\left(  \alpha\right)   &  =-R\left(
\alpha\right)  p_{n-1}^{\left(  j\right)  }\left(  \alpha\right)  -R^{\dag
}\left(  \alpha\right)  p_{n+1}^{\left(  j\right)  }\left(  \alpha\right)
-\delta_{n0}(R_{-}p_{0}^{\left(  j\right)  }\left(  \alpha\right)
+R_{\varepsilon}R_{+}q_{0}^{\left(  j\right)  \ast}\left(  \alpha\right)
),\nonumber\\
\nu q_{n}^{\left(  j\right)  }\left(  \alpha\right)   &  =-R\left(
\alpha\right)  q_{n-1}^{\left(  j\right)  }\left(  \alpha\right)  -R^{\dag
}\left(  \alpha\right)  q_{n+1}^{\left(  j\right)  }\left(  \alpha\right)
-\delta_{n0}(R_{-}q_{0}^{\left(  j\right)  }\left(  \alpha\right)
+R_{\varepsilon}R_{+}p_{0}^{\left(  j\right)  \ast}\left(  \alpha\right)  ).
\label{eq:qn}%
\end{align}
To remove the dependence of $\alpha$, we now let $\tilde{p}_{n}^{\left(
j\right)  }$ $=\left[  R^{n}\left(  \alpha\right)  \right]  ^{\dag}%
p_{n}^{\left(  j\right)  }\left(  \alpha\right)  $ and $\tilde{q}_{n}^{\left(
j\right)  }=\left[  R^{n}\left(  \alpha\right)  \right]  ^{\dag}q_{n}^{\left(
j\right)  }\left(  \alpha\right)  $, which leads to%
\begin{align}
\omega\tilde{p}_{n}^{\left(  j\right)  }  &  =-\tilde{p}_{n-1}^{\left(
j\right)  }-\tilde{p}_{n+1}^{\left(  j\right)  }-\delta_{n0}(R_{-}\tilde
{p}_{0}^{\left(  j\right)  }+R_{\varepsilon}R_{+}\tilde{q}_{0}^{\left(
j\right)  \ast}),\nonumber\\
\nu\tilde{q}_{n}^{\left(  j\right)  }  &  =-\tilde{q}_{n-1}^{\left(  j\right)
}-\tilde{q}_{n+1}^{\left(  j\right)  }-\delta_{n0}(R_{-}\tilde{q}_{0}^{\left(
j\right)  }+R_{\varepsilon}R_{+}\tilde{p}_{0}^{\left(  j\right)  \ast}).
\end{align}
In the above, replacing $n$ with $-n$, we arrive at%
\begin{align}
\omega\tilde{p}_{-n}^{\left(  j\right)  }  &  =-\tilde{p}_{-(n+1)}^{\left(
j\right)  }-\tilde{p}_{-\left(  n-1\right)  }^{\left(  j\right)  }-\delta
_{n0}(R_{-}\tilde{p}_{0}^{\left(  j\right)  }+R_{\varepsilon}R_{+}\tilde
{q}_{0}^{\left(  j\right)  \ast}),\nonumber\\
\nu\tilde{q}_{-n}^{\left(  j\right)  }  &  =-\tilde{q}_{-(n+1)}^{\left(
j\right)  }-\tilde{q}_{-\left(  n-1\right)  }^{\left(  j\right)  }-\delta
_{n0}(R_{-}\tilde{q}_{0}^{\left(  j\right)  }+R_{\varepsilon}R_{+}\tilde
{p}_{0}^{\left(  j\right)  \ast}). \label{eq:pqtild}%
\end{align}
Noting $R^{n}\left(  \alpha\right)  =e\left(  -i\sigma_{y}n\alpha\right)
=R^{-n}\left(  -\alpha\right)  $, we can obtain $p_{-n}^{\left(  j\right)
}\left(  -\alpha\right)  =R^{n}\left(  \alpha\right)  \tilde{p}_{-n}^{\left(
j\right)  }$ and $q_{-n}^{\left(  j\right)  }\left(  -\alpha\right)
=R^{n}\left(  \alpha\right)  \tilde{q}_{-n}^{\left(  j\right)  }$. Then,
multiplying Eq.~(\ref{eq:pqtild}) with $R^{n}\left(  \alpha\right)  $, we
find
\begin{align}
\omega p_{-n}^{\left(  j\right)  }\left(  -\alpha\right)   &  =-p_{-\left(
n+1\right)  }^{\left(  j\right)  }\left(  -\alpha\right)  -p_{-\left(
n-1\right)  }^{\left(  j\right)  }\left(  -\alpha\right)  -\delta_{n0}\left(
R_{-}\tilde{p}_{0}^{\left(  j\right)  }+R_{\varepsilon}R_{+}\tilde{q}%
_{0}^{\left(  j\right)  \ast}\right)  ,\nonumber\\
\nu q_{-n}^{\left(  j\right)  }\left(  -\alpha\right)   &  =-q_{-(n+1)}%
^{\left(  j\right)  }\left(  -\alpha\right)  -q_{-\left(  n-1\right)
}^{\left(  j\right)  }\left(  -\alpha\right)  -\delta_{n0}\left(  R_{-}%
\tilde{q}_{0}^{\left(  j\right)  }+R_{\varepsilon}R_{+}\tilde{p}_{0}^{\left(
j\right)  \ast}\right)  ,
\end{align}
which means $p_{n}=p_{-n}^{\left(  j\right)  }\left(  -\alpha\right)  $ and
$q_{n}=q_{-n}^{\left(  j\right)  }\left(  -\alpha\right)  $ also constitute
the solution of the coupled equation. This solution is equivalent to
$p_{n}=\mathcal{P}_{12}\mathcal{P}_{34}p_{n}^{\left(  j\right)  }\left(
\alpha\right)  $ and $q_{n}=\mathcal{P}_{12}\mathcal{P}_{34}q_{n}^{\left(
j\right)  }\left(  \alpha\right)  $, since they share the same incident wave.
Here, $\mathcal{P}_{j_{1}j_{2}}$ designates the permutation operator acting on
indices $j$, e.g., $\mathcal{P}_{12}\mathcal{P}_{34}p_{n}^{\left(  1\right)
}\left(  \alpha\right)  =p_{n}^{\left(  2\right)  }\left(  \alpha\right)  $
and $\mathcal{P}_{12}\mathcal{P}_{34}p_{n}^{\left(  3\right)  }\left(
\alpha\right)  =p_{n}^{\left(  4\right)  }\left(  \alpha\right)  $. By
comparing the terms between $p_{-n}^{\left(  j\right)  }\left(  -\alpha
\right)  $ and $\mathcal{P}_{12}\mathcal{P}_{34}p_{n}^{\left(  j\right)
}\left(  \alpha\right)  $, we obtain the identity $S_{j^{\prime}j}%
=\mathcal{P}_{12}\mathcal{P}_{34}S_{j^{\prime}j}$, e.g., $S_{12}=S_{21}$, a
straightforward proof of isotropic transport. For example, by specifying
$j=1$, we have
\begin{align}
p_{-n}^{\left(  1\right)  }\left(  -\alpha\right)   &  =l_{n}^{(2)}%
\theta(n-1)+(S_{21}l_{n}^{\left(  1\right)  }+S_{41}l_{n}^{\left(  3\right)
})\theta(n-1)+(S_{11}l_{n}^{\left(  2\right)  }+S_{31}l_{n}^{\left(  4\right)
})\theta(-n),\nonumber\\
\mathcal{P}_{12}\mathcal{P}_{34}p_{n}^{\left(  1\right)  }\left(
\alpha\right)   &  =p_{n}^{\left(  2\right)  }\left(  \alpha\right)
=l_{n}^{(2)}\theta(n)+(S_{12}l_{n}^{\left(  1\right)  }+S_{32}l_{n}^{\left(
3\right)  })\theta(n)+(S_{22}l_{n}^{\left(  2\right)  }+S_{42}l_{n}^{\left(
4\right)  })\theta(-n-1).
\end{align}
The relation $p_{-n}^{\left(  1\right)  }\left(  -\alpha\right)
\equiv\mathcal{P}_{12}\mathcal{P}_{34}p_{n}^{\left(  j\right)  }\left(
\alpha\right)  $ can naturally yield $S_{12}=S_{21}$, $S_{31}=S_{42}$. To this
end, we are convinced to only investigate the scenarios with incoming waves
from negative lattice sites, i.e., $L_{n}^{\left(  j\right)  }$ with $j=1,3$.

\section{Intermediate parameters in the scattering coefficients}

In the expressions of the scattering coefficients $S_{11}$, $S_{33}$, $S_{31}%
$, and $S_{13}$, we have used the intermediate parameters $X$ and $Y$, which
take the following forms,%
\begin{align}
X  &  =\left(  2+\lambda\right)  +\frac{\left(  \lambda^{2}+1\right)
\mu-\lambda^{3}-\lambda-2}{\left(  \mu-\lambda-2\right)  ^{2}-\lambda^{2}%
}=\left(  2+\lambda\right)  +\frac{\left(  \lambda^{2}+1\right)  \mu
-\lambda^{3}-\lambda-2}{\left(  \mu-2\right)  \left(  \mu-2-2\lambda\right)
},\label{eq:X}\\
Y  &  =\lambda\left[  1+\frac{2\mu+\lambda^{2}-2\lambda-3}{\left(  \mu
-\lambda-2\right)  ^{2}-\lambda^{2}}\right]  =\lambda\left[  1+\frac
{2\mu+\lambda^{2}-2\lambda-3}{\left(  \mu-2\right)  \left(  \mu-2-2\lambda
\right)  }\right]  , \label{eq:Y}%
\end{align}
where $\mu=\sqrt{\left(  2\Omega-\omega\right)  ^{2}-4}/g$.

\section{The phenomena in the transport}

\subsection{Distinguishing the spin-reciprocal and spin-nonreciprocal
transport}

\begin{figure}[h]
\includegraphics[width=0.166\textwidth, clip]{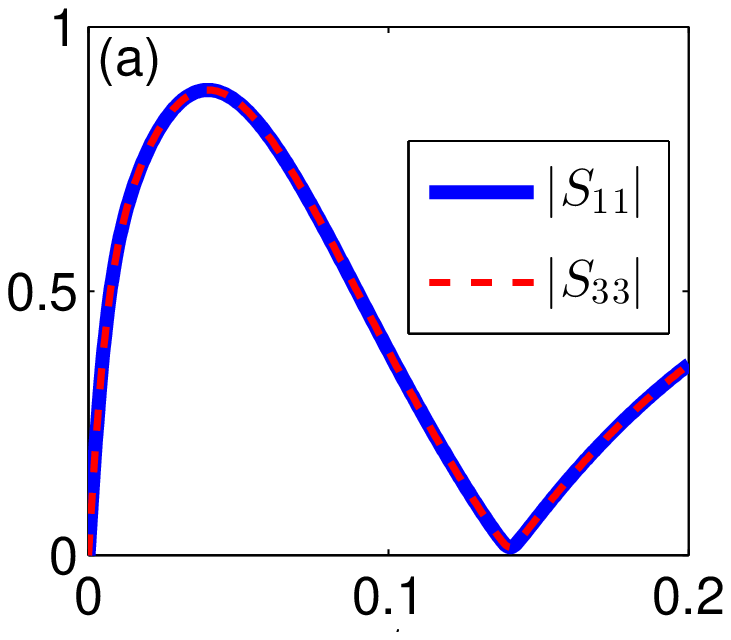}\includegraphics[width=0.166\textwidth, clip]{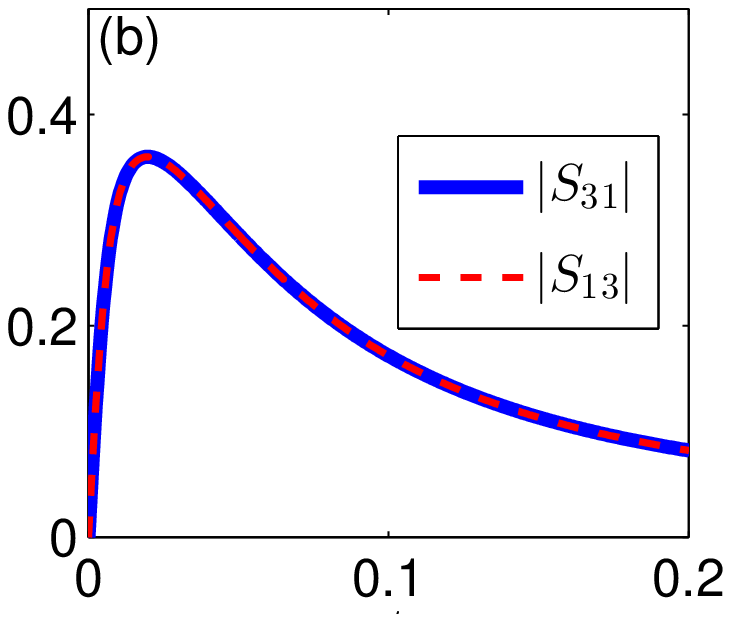}\includegraphics[width=0.166\textwidth, clip]{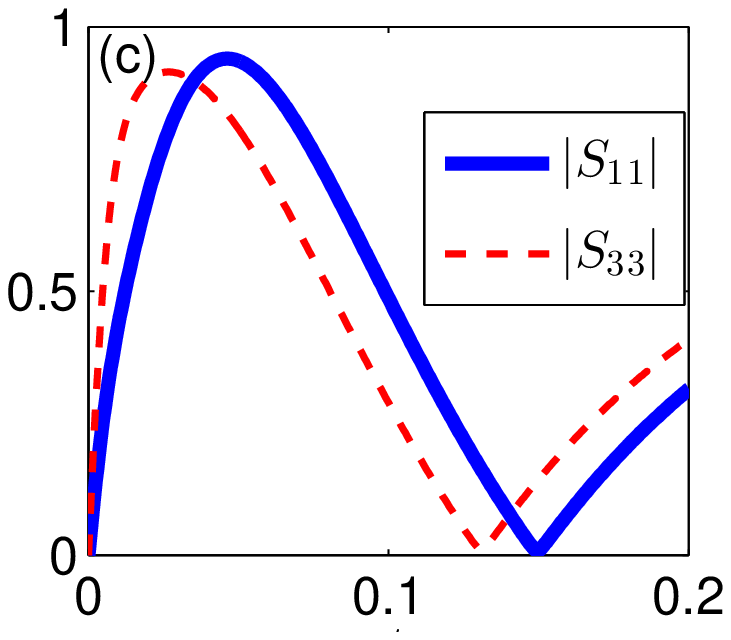}\includegraphics[width=0.166\textwidth, clip]{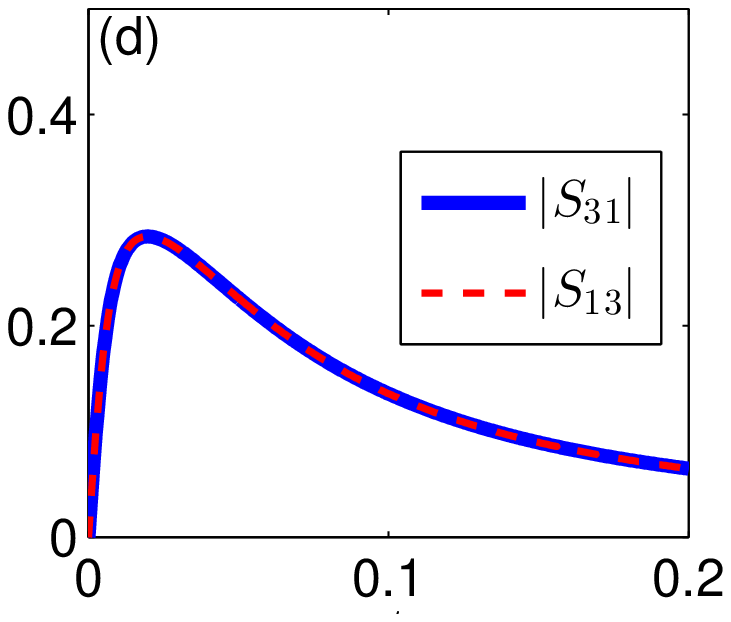}
\caption{(color online) Transmission coefficients ($S_{11}$ and $S_{33}$) and
conversion coefficients ($S_{13}$ and $S_{31}$) plotted against $\varphi/\pi$.
We specify the parameters $g=0.69$, $\lambda=0.1$, and $a=b=\pi/4$. In (a) and
(b), $\varepsilon=\arctan$($\tan a\sin b$), which guarantees spin-reciprocal
transport ($C_{Y}=0$), e.g., $\left\vert S_{11}\right\vert =\left\vert
S_{33}\right\vert $ and $\left\vert S_{31}\right\vert =\left\vert
S_{13}\right\vert $. In (c) and (d), however, $\varepsilon=\arctan$($\tan
a\tan b$)$-\pi/4$, which induces spin-nonreciprocal transport ($C_{Y}%
=-0.6124\neq0$), e.g., $\left\vert S_{11}\right\vert \neq\left\vert
S_{33}\right\vert $. }%
\label{fig:reciprocal&non-reciprocal}%
\end{figure}

Here we will discuss the spin-nonreciprocal / spin-reciprocal transport,
which, differently from the conventional noreciprocal / reciprocal transport
describing spatial unidirectional~\cite{Bai2016PRE,Xu2017PRA} / isotropic
transport, means the transport properties are dependent on / independent of
the spin orientation of incident waves. Such transport processes can be
investigated through the scattering coefficients. Recall that $\mathbf{s}%
_{+,n}$ ($\mathbf{s}_{-,n}$) represents the spin orientation of $l_{n}%
^{\left(  1\right)  }$ [$l_{n}^{\left(  3\right)  }]$. We can find the
spin-reciprocal transport ($\left\vert S_{11}\right\vert =\left\vert
S_{33}\right\vert $) is achieved when $C_{Y}=0$ [or rather, $\tan
\varepsilon=\tan a\sin b$, see Figs.~\ref{fig:reciprocal&non-reciprocal}(a)
and \ref{fig:reciprocal&non-reciprocal}(b)], while the spin-nonreciprocal
transport ($\left\vert S_{11}\right\vert \neq\left\vert S_{33}\right\vert $)
can be achieved once if $C_{Y}\neq0$ [or rather, $\tan\varepsilon\neq\tan
a\sin b$, see Figs.~\ref{fig:reciprocal&non-reciprocal}(c) and
\ref{fig:reciprocal&non-reciprocal}(d)], although $\left\vert S_{31}%
\right\vert =\left\vert S_{13}\right\vert $ always holds.

\subsection{Transparency and blockade}

We now explore the conditions for transparency of the incident wave. In this
case of $j=1$, the transmission coefficients for the incident wave is
\begin{equation}
S_{11}=\frac{i\tilde{\varphi}\left(  i\tilde{\varphi}+X+YC_{Y}\right)
}{\left(  i\tilde{\varphi}+X\right)  ^{2}-Y^{2}}=\frac{i\tilde{\varphi}\left(
i\tilde{\varphi}+X+YC_{Y}\right)  }{\left(  i\tilde{\varphi}+X+Y\right)
\left(  i\tilde{\varphi}+X-Y\right)  }. \label{eq:S11_supp}%
\end{equation}
On the other hand, we note $C_{Y}=\sin\varepsilon\cos a-\cos\varepsilon\sin
a\sin b=\sqrt{\cos^{2}a+\sin^{2}a\sin^{2}b}Y\sin\left(  \varepsilon-c\right)
\in\left[  -1,1\right]  $, where $c=\arctan\left(  \sin b\tan a\right)  $.
Thus, we furthermore have $\left\vert X+YC_{Y}\right\vert \leq\left\vert
X+Y\right\vert $ or $\left\vert X+YC_{Y}\right\vert \leq\left\vert
X-Y\right\vert $, which leads to
\begin{equation}
\left\vert S_{11}\right\vert =\left\vert \frac{i\tilde{\varphi}}{\left(
i\tilde{\varphi}+X+Y\right)  }\right\vert \left\vert \frac{i\tilde{\varphi
}+X+YC_{Y}}{i\tilde{\varphi}+X-Y}\right\vert =\left\vert \frac{i\tilde
{\varphi}+X+YC_{Y}}{\left(  i\tilde{\varphi}+X+Y\right)  }\right\vert
\left\vert \frac{i\tilde{\varphi}}{i\tilde{\varphi}+X-Y}\right\vert \leq1.
\end{equation}
Here, the \textquotedblleft=\textquotedblright\ sign is achieved when
\[
C_{Y}=\mp1,X\pm Y=0.
\]
In this condition, we actually have $S_{11}=1$, meaning the incident wave
$L_{n}^{\left(  1\right)  }$ is transparent. In the case of $j=3$, the
transmission coefficients for the incident wave is%
\begin{equation}
S_{33}=\frac{i\tilde{\varphi}\left(  i\tilde{\varphi}+X-YC_{Y}\right)
}{\left(  i\tilde{\varphi}+X\right)  ^{2}-Y^{2}}=\frac{i\tilde{\varphi}\left(
i\tilde{\varphi}+X-YC_{Y}\right)  }{\left(  i\tilde{\varphi}+X+Y\right)
\left(  i\tilde{\varphi}+X-Y\right)  }. \label{eq:S33_supp}%
\end{equation}
We can similarly obtain the transparency ($S_{33}=1$) condition is
\begin{equation}
C_{Y}=\mp1,X\mp Y=0.
\end{equation}
Moreover, $X+Y=0$ yields T1: $\mu=-\frac{1}{2}\left(  \lambda-3\right)
\left(  \lambda+1\right)  $, while $X-Y=0$ yields T2: $\mu=\frac{3}{2}\left(
\lambda+1\right)  .$

We now explore the blockade ($S_{jj}=0$) condition for the incident wave
$L_{n}^{\left(  j\right)  }$. In the denominator of Eq.~(\ref{eq:S11_supp}),
$X\pm Y$ can be transformed into
\begin{align}
X+Y  &  =\frac{\left(  \lambda+1\right)  \left(  2\mu-3\lambda-3\right)  }%
{\mu-2\lambda-2},\\
X-Y  &  =\frac{2\mu+\lambda^{2}-2\lambda-3}{\mu-2}.
\end{align}
The blockade can only occur at B1: $\mu=2\lambda+2=0$ (making $X+Y=\infty)$,
or B2: $\mu=2$ (making $X-Y=\infty$), when the numerator of $S_{11}$ and
$S_{33}$ should be kept limited. In detail, if the blockade point is
$\mu-2\lambda-2$ ($\mu=2$), to make $S_{11}=0$, there should be $\lim
_{\mu\rightarrow2\lambda+2\text{ (}\mu\rightarrow2\text{)}}\left(
X+YC_{Y}\right)  =$ \textit{Constant}, which yields $C_{Y}=-1$ ($C_{Y}=1$).
Similarly, if the blockade point is $\mu-2\lambda-2$ ($\mu=2$), to make
$S_{33}=0$ [see Eq.~(\ref{eq:S33_supp})], there should be $\lim_{\mu
\rightarrow2\lambda+2\text{ (}\mu\rightarrow2\text{)}}\left(  X-YC_{Y}\right)
=$ \textit{Constant}, which yields $C_{Y}=1$ ($C_{Y}=-1$). Note that when
$C_{Y}=\pm1$, the transmission coefficients are reduced to an analog form of
the spinless case~\cite{Vicencio2007PRL}.

\begin{figure}[h]
\includegraphics[width=0.155\textwidth, clip]{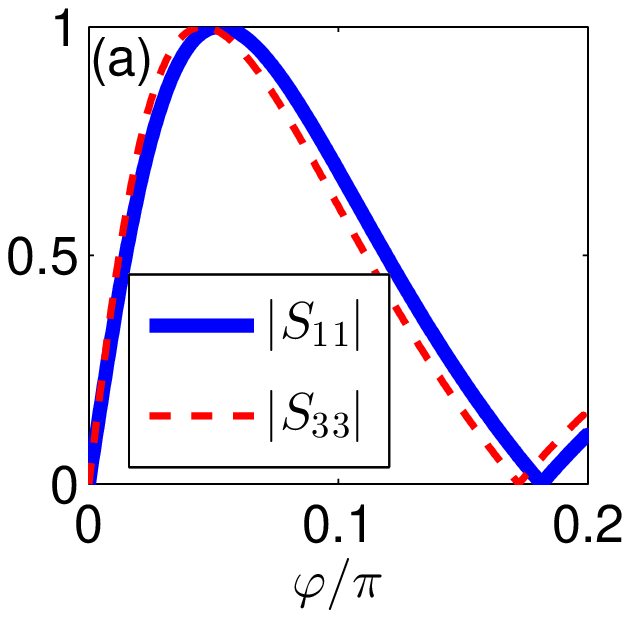}
\includegraphics[width=0.155\textwidth, clip]{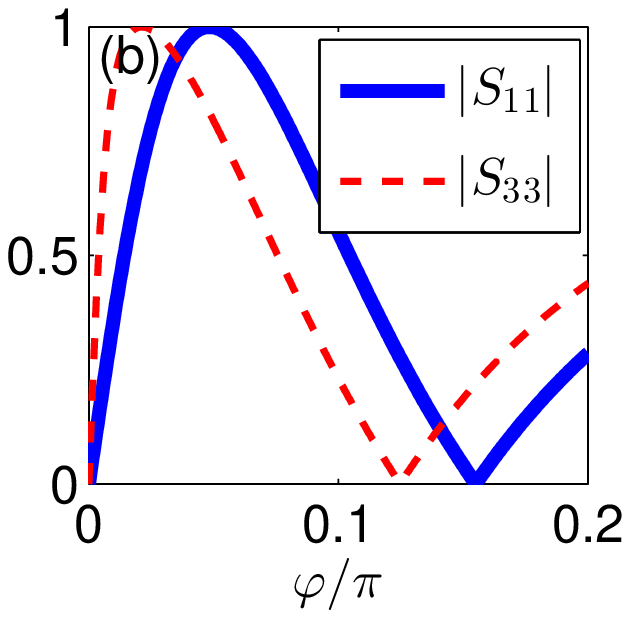}
\includegraphics[width=0.155\textwidth, clip]{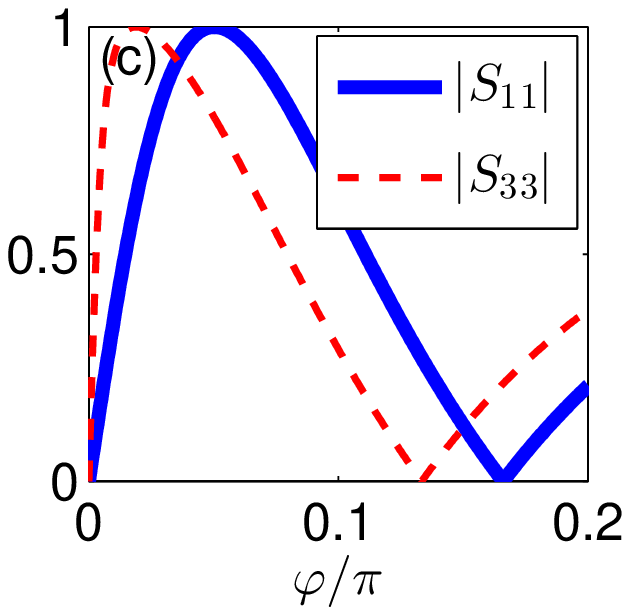}
\includegraphics[width=0.155\textwidth, clip]{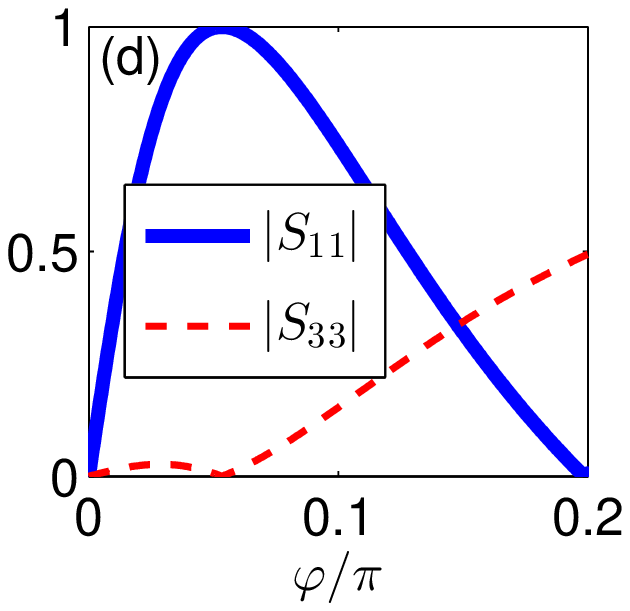} \caption{(color
online) Transmission coefficients $S_{11}$ and $S_{33}$ for spin-nonreciprocal
transport ($\left\vert S_{11}\right\vert \neq\left\vert S_{33}\right\vert $)
at (a) $g=0.9$, $\lambda=0.025$, (b) $g=0.69$, $\lambda=0.1$, (c) $g=0.75$,
$\lambda=0.1$, and (d) $g=0.7788$, $\lambda=0.3333$, where $\varepsilon
=a-\pi/2$ such that $\mathbf{s}_{\varepsilon,n}$ orients parallel to
$\mathbf{s}_{+,n}$ and $\left\vert S_{31}\right\vert =\left\vert
S_{13}\right\vert =0$ (not shown). }%
\label{fig:T&B&ISO}%
\end{figure}

Now we summarize the results. When $C_{Y}=-1$, we achieve $S_{11}=1$ and
$S_{11}=0$ respectively at
\begin{equation}
\text{T1: }\mu=-\frac{1}{2}\left(  \lambda-3\right)  \left(  \lambda+1\right)
\text{ and B1: }\mu=2\lambda+2=0,
\end{equation}
but achieve $S_{33}=1$ and $S_{33}=0$ respectively at%
\begin{equation}
\text{T2: }\mu=\frac{3}{2}\left(  \lambda+1\right)  \text{ and B2: }\mu=2.
\end{equation}
By comparison, when $C_{Y}=1$, we achieve $S_{11}=1$ and $S_{11}=0$
respectively at T2 and B2, but achieve $S_{33}=1$ and $S_{33}=0$ respectively
at T1 and B1. Since we have assumed $b\in\left[  0,\pi\right]  $, then the
condition $C_{Y}=\mp1$ yields $b=\pi/2$, and $\varepsilon-a=\mp\pi/2$. Here,
$b=\pi/2$ leads to $\mathbf{s}_{\pm,n}=\pm2[\sin a\cos\left(  2n\alpha\right)
\mathbf{e}_{x}+\cos a\mathbf{e}_{y}-\sin a\sin\left(  2n\alpha\right)
\mathbf{e}_{z}]$, meaning $\mathbf{s}_{\pm,0}$ orients within the $xoy$ plane
[note that $\mathbf{s}_{\pm,n}$ represent the spin orientation of
$l_{n}^{\left(  1\right)  }$ and $l_{n}^{\left(  3\right)  }$]. Furthermore,
$\varepsilon-a=\mp\pi/2$ leads to $\mathbf{s}_{\varepsilon,n}=$ $\pm2\left(
g/\gamma\right)  \kappa^{2\left\vert n\right\vert }[\sin a\cos\left(
2n\alpha\right)  \mathbf{e}_{x}+\cos a\mathbf{e}_{y}-\sin a\sin\left(
2n\alpha\right)  \mathbf{e}_{z}]$, i.e., $\mathbf{s}_{\varepsilon,n}=$
$\left(  g/\gamma\right)  \kappa^{2\left\vert n\right\vert }\mathbf{s}_{\pm
,n}$, meaning the strong localized mode $d_{n}$ should orient identical to
$l_{n}^{\left(  1\right)  }$ or $l_{n}^{\left(  3\right)  }$. Here, T1, T2, B1
and B2 determine the energy $\omega$ at the transparency or blockade points.
Moreover, $C_{Y}=\mp1$ leads to $S_{31}=S_{13}=0$, signifying no conversion
between $l_{n}^{\left(  1\right)  }$ and $l_{n}^{\left(  3\right)  }$ in the
output fields. In Fig.~\ref{fig:T&B&ISO}, we have shown $\left\vert
S_{11}\right\vert $ and $\left\vert S_{33}\right\vert $ for different
parameters. 

\subsection{Spin conversion}

We now focus on the spin conversion. Our mission is to find the maximum
conversion efficiency. Since $\left\vert S_{31}\right\vert =$ $\left\vert
S_{13}\right\vert $, we only need to investigate:%
\begin{equation}
S_{31}=\frac{i\tilde{\varphi}\left(  iY\right)  \left(  ie^{ib}\sin
a\sin\varepsilon-C_{\varepsilon}\cos\varepsilon\right)  }{\left(
i\tilde{\varphi}+X\right)  ^{2}-Y^{2}}%
\end{equation}
where $C_{\varepsilon}=\cos^{2}\left(  a/2\right)  +e^{i2b}\sin^{2}\left(
a/2\right)  $.

We define $M=ie^{ib}\sin a\sin\varepsilon-C_{\varepsilon}\cos\varepsilon,$
which is independent of $\tilde{\varphi}=2g^{-1}\sin\varphi$. To examine the
maximum value of $M$, we have
\begin{align}
\left\vert M\right\vert ^{2}  &  =\frac{\left(  1+\sin^{2}a\cos^{2}b\right)
+\left(  \cos^{2}a-\sin^{2}a\sin^{2}b\right)  \cos2\varepsilon+\sin2a\sin
b\sin2\varepsilon}{2}\nonumber\\
&  =\frac{\left(  1+\sin^{2}a\cos^{2}b\right)  +\left(  \cos^{2}a+\sin
^{2}a\sin^{2}b\right)  \cos\left(  2\varepsilon-2\theta^{\prime}\right)  }{2}%
\end{align}
Here, $\theta^{\prime}$ is determined by $\tan2\theta^{\prime}=\sin2a\sin
b/\left(  \cos^{2}a-\sin^{2}a\sin^{2}b\right)  $, which can be simplified into
$\tan\theta^{\prime}=\tan a\sin b$. First, it is obvious that $\left\vert
M\right\vert ^{2}\geq\sin^{2}a\cos^{2}b$, where the lower bound is achieved
when $\varepsilon=\theta^{\prime}\pm\pi/2$. In particular, if $b=\pi/2$, the
lower bound becomes zero with $\varepsilon=a\pm\pi/2$, which is the very
condition for transparency and blockade. Similarly, we also have $\left\vert
M\right\vert ^{2}\leq1$, where the upper bound is achieved when $\varepsilon
=\theta^{\prime}$ or $\varepsilon=\theta^{\prime}+\pi$, i.e., $\tan
\varepsilon=\tan a\sin b$. This condition makes $C_{Y}=0$, meaning the
transport is spin-reciprocal.

\begin{figure}[h]
\includegraphics[width=0.155\textwidth, clip]{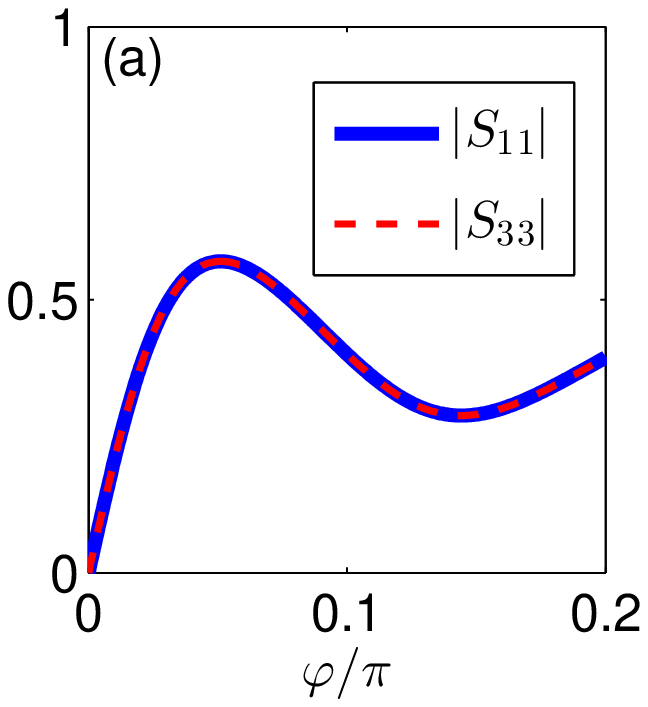}
\includegraphics[width=0.155\textwidth, clip]{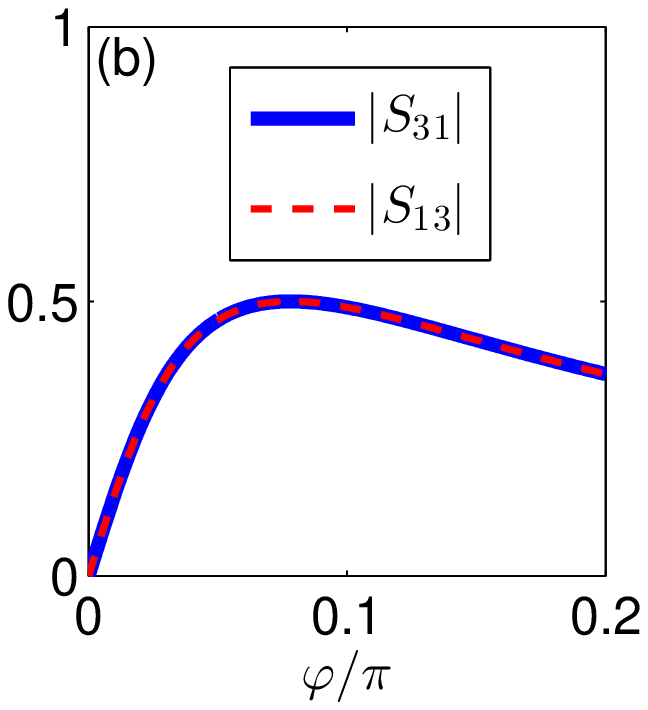}\caption{(color
online) (a) Transmission coefficients $S_{11}$ and $S_{33}$, and conversion
efficiencies $S_{31}$ and $S_{13}$ for spin-reciprocal transport ($\left\vert
S_{11}\right\vert =\left\vert S_{33}\right\vert $) at localization grade
$g=0.5$ and interaction ratio $\lambda=1$. Besides, we specify $\varepsilon
=\arctan(\tan a\sin b)$, which favors the maximization of $S_{31}$ and
$S_{13}$.}%
\label{fig:conv}%
\end{figure}

Suppose $\left\vert M\right\vert =1$ is achieved by setting $\tan
\varepsilon=\tan a\sin b$, we now continue to maximize $\left\vert
S_{31}\right\vert $ via examining $\tilde{\varphi}$ (see Fig.~\ref{fig:conv}).
In this manner, we can easily obtain%
\begin{equation}
\left\vert S_{31}\right\vert ^{2}=\frac{\tilde{\varphi}^{2}Y^{2}}%
{[\tilde{\varphi}^{2}+\left(  X+Y\right)  ^{2}][\tilde{\varphi}^{2}+\left(
X-Y\right)  ^{2}]}\leq\frac{1}{4},
\end{equation}
where the \textquotedblleft=\textquotedblright\ sign is achieved at
$\tilde{\varphi}^{2}=Y^{2}-X^{2}$ [i.e., $4-\omega^{2}=g^{2}(Y^{2}-X^{2})]$,
and the maximum $\left\vert S_{31}\right\vert $ is $1/2$. We can easily verify
that the condition $\tilde{\varphi}^{2}=Y^{2}-X^{2}$ and $\tan\varepsilon=\tan
a\sin b$ also leads to $S_{11}=S_{33}=1/2$. 

\bibliographystyle{apsrev}
\bibliography{SOT-references}